\documentclass[floatfix,showkeys,twocolumn,citeautoscript,showpacs,amsmath,amssymb,epsf,aps]{revtex4}
\usepackage{dcolumn}
\usepackage{epsfig}

\usepackage{graphicx}
\usepackage{longtable}
\usepackage{dcolumn}
\usepackage{bm}

\def\ocm{ cm$^{-1}$ }
\def\cm{ cm$^{-1}$ }

\def\siA{Si$_{10}$H$_{16}$}
\def\SiA{Si$_{10}$H$_{16}$}
\def\silyA{Si$_{14}$C$_{24}$H$_{72}$}
\def\SilyA{Si$_{14}$C$_{24}$H$_{72}$}
\def\SilyA{C$_{24}$Si$_{14}$H$_{72}$}
\def\silyA{C$_{24}$Si$_{14}$H$_{72}$}

\begin{document}
\preprint{USC/002}
\title{ Equilibrium structure  and  vibrational spectra of  sila-adamantane} 

\author{Rajendra R. Zope$^{1,2}$, Tunna Baruah$^{2}$, Mark R. Pederson$^{3}$, and S. L. Richardson$^{1,3}$}

\affiliation{$^1$NSF CREST Center for Nanomaterials Characterization Science and Process Technology \\
Howard University,
School of Engineering, 
2300 Sixth Street, N.W. 
Washington, D.C. 20059}

\affiliation{$^2$Department of Physics, The University of Texas at El Paso, El Paso, TX 79958 }

\affiliation{$^3$Center for Computational Materials Science,
   Code 6392, Naval Research Laboratory, Washington, DC 20375}


%
%
%
%
%



\date{October 27, 2006}

%

\begin{abstract}
The recent synthesis of a four-fold silylated sila-adamantane molecule (C$_{24}$H$_{72}$Si$_{14}$, T$_d$),
[J. Fischer, J. Baumgartner, and C. Marschner, {\it Science} {\bf 310,} (2005) 825] is the first attempt 
of making the silicon analogue of adamantane. It  has adamantane-like Si$_{10}$ core, capped 
by methyl and sily-methyl ligands. We report its electronic structure, vibrational spectrum, and 
the infra-red and Raman spectra calculated within the  density functional formalism  
using large polarized Gaussian basis sets.  The properties of sila-adamantane 
are compared with exact silicon analogue (Si$_{10}$H$_{14}$) of adamantane.  Results show that 
replacing hydrogens in \SiA\, by methyl and silymethyl ligands results in expansion of the 
Si$_{10}$ core and are responsible for large number of peaks  in the Raman 
spectrum.   The Si-C stretch at 664 \cm and methyl deformations frequencies compare well with recent 
measurements of  vibrational frequencies of methylated silicon surface. 


\end{abstract}

\pacs{ }

\keywords{ adamantane, sila-adamantane, diamondoid}

\maketitle

 \section{Introduction}
      Silicon is of considerable interest in nanotechnology and in semiconductor industry. 
In bulk phase,  silicon has diamond lattice like carbon.  The diamond structure can 
be built by  translating the smallest unit of the carbon lattice called adamantane.
The adamantane has 10 carbon atoms that are capped by hydrogen atoms that 
passivate the dangling bonds of  carbon atoms in the sp$^3$ bonding.
The adamantane and a few low-order diamondoids have been synthesized\cite{adamantane}.
 Recently, isolation of several new lower- and medium-order diamondoids from petroleum oil 
was reported  by the Molecular Diamond Technologies Group at Chevron-Texaco\cite{Cheveron}.
This has led to a  renewal of interest  in the chemistry and physics of the diamondoids\cite{Diamondoids}.

\begin{figure}
\epsfig{file=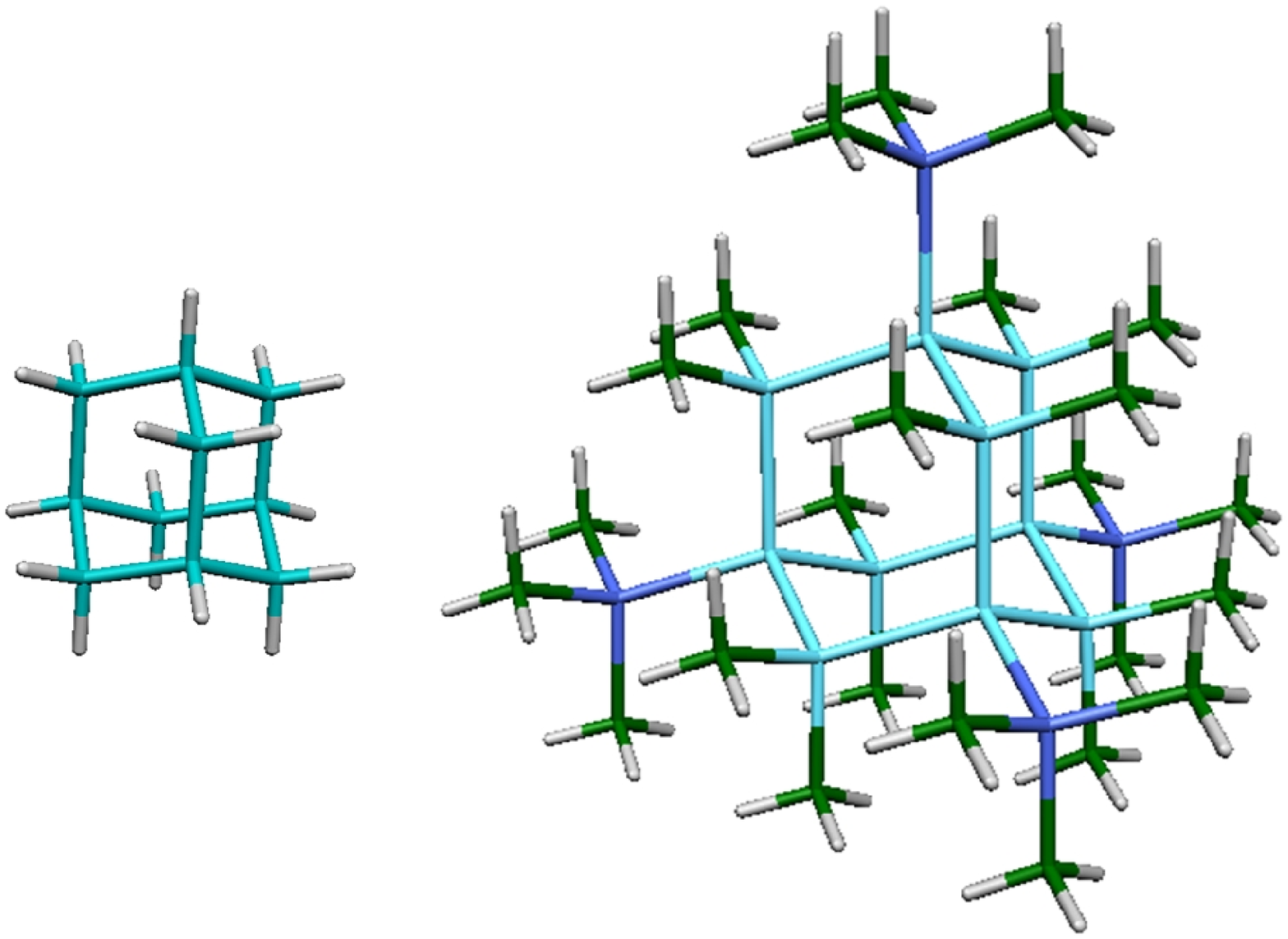,width=8.5cm,clip=true}
\caption{\label{fig:str} (Color online) The optimized structures of 
sila-adamantane \SiA\, and silylated sila-adamantane \SilyA.}
\label{fig1}
\end{figure}

   The search for silicon analogue of adamantane, called sila-adamantane  \SiA,
has so far been unsuccessful. There has been however some progress.  A recent report 
by Ficher, Baumgartner and Marschner (FBM) of 
synthesis  of silylated sila-adamantane is a 
breakthrough in the synthesis of sila-adamantane\cite{Expt}. 
Instead of hydrogen atoms as capping elements as in the exact silicon 
analogue of adamantane, silicon atoms in sila-adamantane are capped by methyl 
and trimethylsilyl groups (Cf. Fig. ~\ref{fig:str}).  It was characterized using the 
nuclear magnetic resonance (NMR) spectroscopy.  The ultra-violet and visible spectra 
were also measured which showed strong absorption at 222 nm (5.58 eV). 
Followed by this report, Piccheri  reported electronic structure of \SilyA\ within 
Perdew-Burke-Ernzerhof density functional model. He reported structural 
parameters calculated using the 3-21G* basis sets and density fitting 
methodology\cite{Pichierri}.

          In the present article, we report the electronic structure of 
sila-adamantane examined within the density functional theory 
using large polarized Gaussian basis sets. Furthermore, at the same level of theory 
we calculate 
vibrational frequencies, the infra-red and Raman spectra.
 We also calculate these properties for exact silicon analogue 
of adamantane (Si$_{10}$H$_{16}$), and compare them with those of sila-adamantane 
\silyA\, to examine the effect of methyl and silymethyl ligands.

\section{Formalism}
      
         Our calculations are performed within the Kohn-Sham formulation\cite{KS}
of the density functional theory\cite{HK} using NRLMOL code\cite{NRLMOL}. The 
molecular orbitals are expressed as linear combination of Gaussian 
functions.  The exchange-correlation effects are described within the 
generalized gradient approximation using Perdew-Burke-Ernzerhof\cite{PBE} 
parameterization.  For a reliable prediction of properties, we use large 
polarized basis sets that are specially optimized for the PBE 
functional\cite{Porezag99}. 
Thus the basis for Si contains 6 s-, 5 p-, and 3 d-type Gaussians each 
contracted from 16 primitive functions. For C, 12 primitive functions were used to
contract 5 s-, 4p-, and 3 d-type basis functions. 
The basis for H consist of 4 s-, 3 p-, and 1 d- type basis functions.
In total 2614 basis 
functions were used in calculation of \SilyA. Additional diffuse functions were used for the 
calculations of polarizabilities.
The exchange-correlation contribution to matrix elements and total energy 
are obtained numerically using efficient numerical grids employed 
in NRLMOL\cite{NRLMOL}.  The vibrational frequencies were obtained by  
diagonalizing the dynamical matrix built from the forces computed at 
perturbed geometries.
Finally, the infra-red and Raman spectra were computed from the dipole 
moment and polarizability derivatives, respectively.
 
\begin{figure}
\epsfig{file=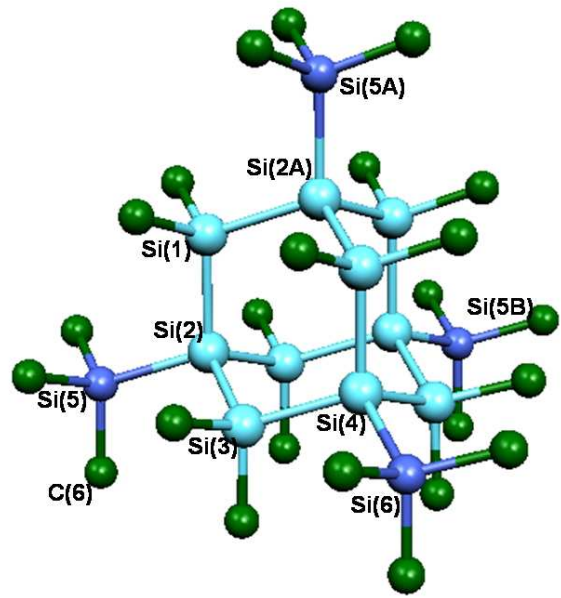,width=8.5cm,clip=true}
\caption{\label{fig:big} (Color online) The optimized structure of \SilyA\,
without hydrogens. The Si atoms that are part of the silymethyl 
ligands are shown in dark blue for clarity. The atomic labels are identical 
to those in experimental structure from Ref. \onlinecite{Expt}.  }
\end{figure}

\section{Results}
     
     The  optimized structure of \SiA\, and \SilyA\, are shown in Fig. 1.
The \SilyA\,  is also shown without hydrogens in Fig. 2.  To distinguish 
the Si atoms that belong to silymethyl-ligand, these four atoms are shown 
in dark blue. The \SiA\, core in \SilyA\, is clearly seen.  The atoms 
have labels exactly identical to those given in the supplementary 
information of the experimental study in  Ref. \onlinecite{Expt}.
Both structures were optimized within the T$_d$ symmetry. 
 The bond distance between two silicon atoms (Si(1)-Si(2)) that belong to 
the tetrahedral adamantane
core (Si(10)) is 2.39 \AA, in good agreement with experimental value of 2.36 \AA. 
In case of \SiA, this bond length is  calculated to be 2.36 \AA.  The Si-SiCH$_3$ 
bond length (Si(2)-Si(5) distance) is also 2.39 \AA\, while the experimental value is 2.35 \AA.
The bond angle between the Si-Si bonds from the Si$_{10}$ core  (Si(2)-Si(1)-Si(2A)) 
in \SilyA\, is 110$^o$ and is in excellent agreement with the experimental value 
of 110.3$^o$.  Similarly, 
very good agreement is observed for the dihedral angle between three bonds from the Si$_{10}$ 
core ( Si(2A)-Si(1)-Si(2)-Si(3)).  The predicted angle is 59.8$^o$ while  
experimental value is 59.9$^o$.  These bond angles 
are essentially  similar in \SiA\, (angles 110.1$^o$ and 59. 7$^o$). As the bond angles and dihedral 
angles in \SiA\, are essentially similar to those in the Si$_{10}$ core in \silyA, it appears 
that replacing hydrogens by  methyl and trimethylsilyl ligands leads to  an 
expansion of the Si$_{10}$ core.

   \section{Electronic structure properties}
            The  fully occupied highest occupied  molecular orbital (HOMO) of \SilyA\, 
is three fold degenerate 
and has T$_2$ symmetry while its LUMO has 
 A$_1$ symmetry. The HOMO-LUMO (HL) gap is   4.16 eV. For the  \SiA, the HL gap is 4.82 eV.
The HL gaps reported earlier with a smaller 3-21G* orbital basis and the density 
fitting methodology are in the range 4.38-4.53 eV depending on the choice of exchange-correlation 
parameterization\cite{Pichierri}.
We also calculated the HOMO-LUMO gap for the \SiA\, using the time dependent DFT (TDDFT)
and using the Perdew-Wang GGA as implemented in ABINIT\cite{ABINIT}.  The calculated gap using TDDFT
of 4.6 eV is only marginally larger than the PBE-DFT gap.  The optical gaps of \SiA\, have been 
studied using more complex quantum Monte Carlo and GW approximations\cite{benedict}. 
These methods
give value of  9.2 eV for the optical gaps. Our attempts to compute the TDDFT
optical gap for \SilyA\, were not successful due to large memory requirement.

 
\begin{figure}
\epsfig{file=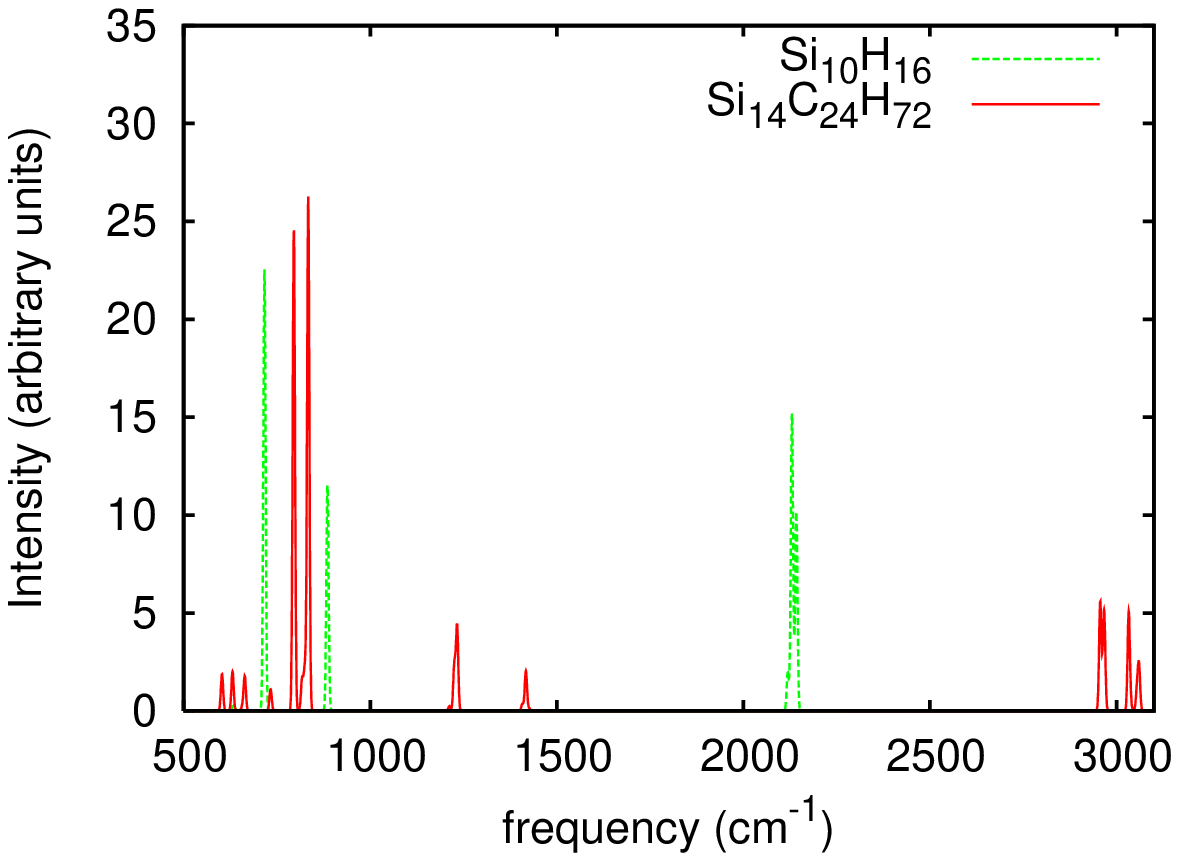,width=8.5cm,clip=true}
\caption{\label{fig:ir} (Color online) The infra-red spectra of \SiA\,  and \SilyA.}
\end{figure}

\begin{figure}
\epsfig{file=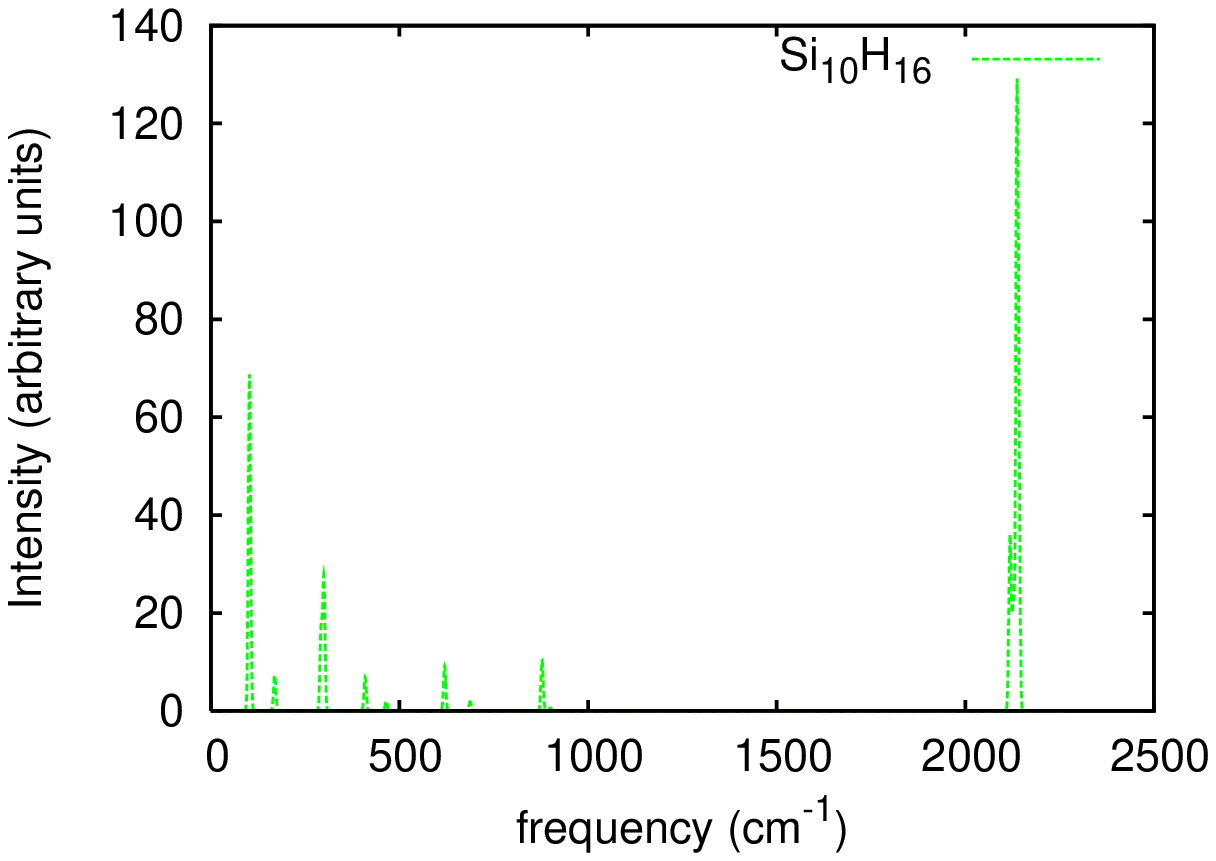,width=8.5cm,clip=true}
\caption{\label{fig:raman_small} (Color online) The Raman spectra of \SiA.}
\end{figure}

\begin{figure}
\epsfig{file=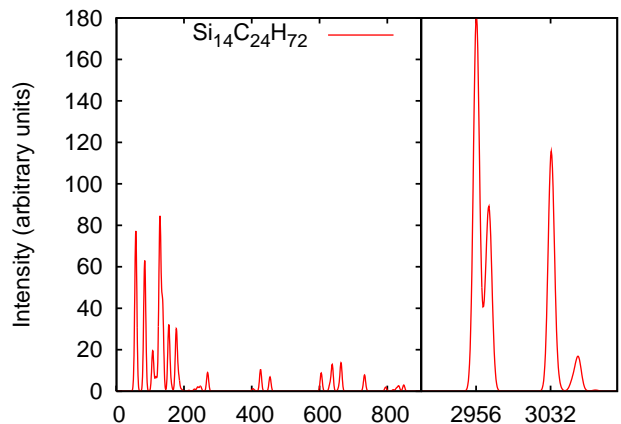,width=8.5cm,clip=true}
\caption{\label{fig:raman_big} (Color online) The Raman spectra of \SilyA. It is split into low frequency 
 region 50-1000\cm  and high frequency 2800-3100 \cm (CH stretches) region.}
\end{figure}

   \section{Vibrational spectra}
  \subsection{\SiA}
   The \SiA\, cage has  $T_d$  point group symmetry. 
It has in total 72 vibrational modes which 
could be classified as  5A$_1$ +  A$_2$ + 6 E + 7T$_1$ + 11 T$_2$.  
Of the total 11 T$_2$ modes that are infra-red active, only four have appreciable 
intensity. 
%
%
The calculated IR spectra is shown in Fig. \ref{fig:ir}. 
The first absorption is at 716 \cm (7.61 D/\AA$^2$) due to the SiH$_2$ wagging. The mode at
885\cm  (3.85  D/\AA$^2$) corresponds to the bending (scissor) mode of SiH$_2$.
The peaks for the two SiH stretching modes are at higher frequencies 
of 2130 \ocm and 2141  \ocm are coalesced in the figure. 
The quantities in bracket are intensities.

   The Raman spectra of \SiA\, is shown in Fig. ~\ref{fig:raman_small}.
There are  5A$_1$, 6 E, and 11 T$_2$  Raman active modes. 
The  fully symmetric A$_1$ mode  at 2137 \ocm is the symmetric SiH  stretch and is the most 
intense (intensity 130 \AA$^4$/amu).  The absorption due to fully symmetric stretch of 
Si-Si bonds (breathing of silicon cage) occurs at lower (due to larger mass) frequency of 299 \cm.
The second most intense absorption is seen at lower 
wave number of 102 \ocm with intensity of 68 \AA$^4$/amu. This is three fold 
degenerate T$_1$ mode due to rocking  motion of SiH$_2$. The 
Raman active  doubly degenerate mode at 103 \cm  is due to the deformation of silicon cage.
The absorptions at larger frequencies ( $> 600$\cm) involve SiH$_2$ deformations and SiH stretches.
Less intense absorption  are  620 \cm (SiH$_2$ twist, intensity 9 \AA$^4$/amu), 688 \cm 
(SiH$_2$ and SiH rocking, intensity 2 \AA$^4$/amu), and 878 \cm (SiH$_2$ scissor, 10 \AA$^4$/amu ).

  \subsection{\SilyA}

 We computed the vibrational frequencies, within the harmonic approximation, 
with different possible orientations of methyl groups that are consistent with the $T_d$ symmetry. 
In all cases we find 9 imaginary frequencies for \SilyA\,  due to the 
rotation  of CH3 group.   These rotations seen in the gas phase will be frozen in the crystalline state.
The total 324 vibrational modes 
in the \SilyA\,  can be classified as 17A$_1$ + 10 A$_2$ + 27 E + 37T$_1$ + 44 T$_2$.  
There are a total of 44 IR active modes.  The calculated infra-red absorption spectrum is displayed 
in Fig. ~\ref{fig:ir}.
The two conspicuous peaks at lower frequencies of 795 \cm (8 D/\AA$^2$) and  834 \cm   ( 9 D/\AA$^2$)
are due to the SiCH$_3$ rocking/wagging motions. These modes are similar to the SiH$_2$ wagging and rocking 
seen in \SiA\.  The weak absorption at 1232 \cm 
( 1.4 D/\AA$^2$) is the Si-CH$_3$ umbrella mode.  The three peaks at higher frequencies 
in the frequency range  2956-3032 \cm with intensities of 1.3-1.4 D/\AA$^2$ are due to the 
symmetric and asymmetric C-H stretches.  A number of vibrational modes results in weak absorption,
with intensities less than 1  D/\AA$^2$. These are  Si-C asymmetric stretch at 664 \cm (0.6  D/\AA$^2$), 
Si-Si-Si deformation at 631 \cm (0.7  D/\AA$^2$), Si-C symmetric stretch at 603 \cm (0.6  D/\AA$^2$),
CH$_3$ deformation at 1417 \cm (0.7  D/\AA$^2$), and C-H symmetric stretch at 3060 (0.8  D/\AA$^2$).
The compilation of the frequencies is given in Table \ref{tab:IR_big}.

\begin{table}[h!]
\caption{\label{tab:IR_big} The IR active frequenters for the \SilyA. The 
 stretch of absorptions is given in bracket with following abbreviations:
 w: weak, vs: very strong, m: medium. The experimental numbers are for methylated 
 silicon(111) surface obtained by high-resolution electron energy loss spectroscopy (HREELS)\cite{Yamada} 
 and transmission infra-red (TIR)\cite{Webb} spectroscopy.}
\begin{tabular}{lccl}
\hline
 Freq. (\cm)    &  HREELS  & TIR    &     Mode description \\
\hline
 664 (w)        &  678   &       &  Si-C asymmetric stretch   \\
 603 (w)        &        &       & Si-C symmetric stretch   \\
 631  (w)       &        &       & Si-Si-Si deformation   \\
 796(vs)        &        &       &  SiCH$_3$ bending    \\
 834(vs)        &        &        & CH$_3$  rocking    \\
 1233(m)        &        &   1257 &    CH$_3$ Umbrella    \\
 1417 (w)       &  1424  &   2856 &    CH$_3$ deformation    \\
 2956 (m)       &        &   2909 &    C-H symmetric stretch    \\
 2969 (m)       &  2916  &  2909 &    C-H symmetric stretch    \\
 3032 (m)       &  2989  &  2928 &    C-H asymmetric stretch    \\
\hline
\end{tabular}
\end{table}

\begin{figure}
\epsfig{file=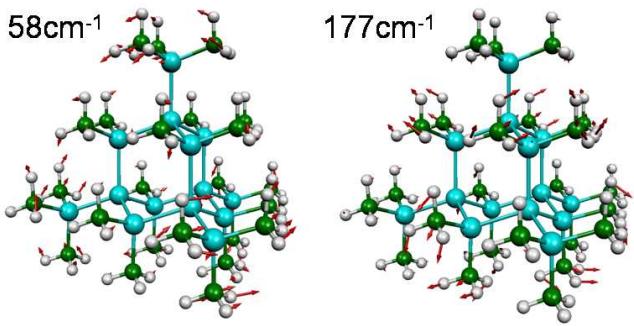,width=8.5cm,clip=true}
\caption{\label{fig:Fig59-177} (Color online) The portrayal of 
vibrational modes:   Left: doubly degenerate mode at 59 \cm, Right: Nondegenerate mode at 177 \cm 
due to breathing motion of Si atoms of Si$_{10}$ core. (See text for more details).}
\end{figure}

\begin{figure}
\epsfig{file=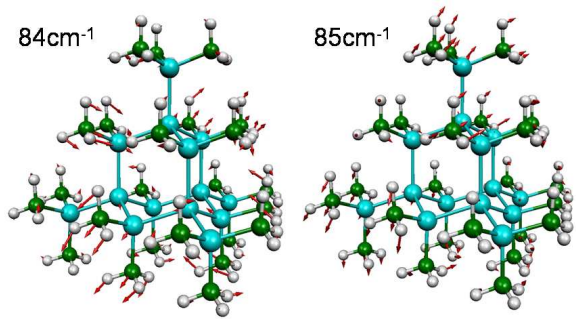,width=8.5cm,clip=true}
\caption{\label{fig:Fig84-85} (Color online) The depiction of Raman active vibrational modes due 
to distortion of Si$_{10}$ core:
Left: doubly degenerate mode at 84 \cm, Right: Triply degenerate mode at 85 \cm.}
\end{figure}

        The Raman spectrum  of \silyA\, is shown in Fig. ~\ref{fig:raman_big}. The spectrum 
can be split into two parts: the absorptions in the low-intermediate frequency range 
(60-200 \cm ) and  intense absorptions in the frequency range 2900-3100 \cm. The doubly 
degenerate  absorption at 59 \cm (77  \AA$^4$/amu) involves rocking like motion of Si-Si(CH$_3$)$_3$.
This mode is depicted in Fig. \ref{fig:Fig59-177}.
The absorption at 84 \cm (63  \AA$^4$/amu) is due to a doubly  degenerate mode that results in 
deformation of Si-Si-Si bonds in the Si$_{10}$ core similar to the mode at 103 \cm seen in \SiA\. 
The triplet degenerate mode also due to the distortion of Si$_{10}$ core at 85 \cm is merged with 
the peak at 84 \cm. These two modes are shown in Fig. \ref{fig:Fig84-85}.
The weak absorptions at 108,  155 \cm are due to 
methyl rotations. The most intense peak in the low frequency region at 130 \cm (85  \AA$^4$/amu) 
is due to non-degenerate A$_1$ mode that results in breathing motion of Si atoms to which Si(CH$_3$)$_3$ 
ligands are attached. The breathing motions of rest of Si atoms in Si$_{10}$ core (to which two 
CH$_3$ groups are attached) leads to an absorption at 177 \cm ( \AA$^4$/amu).  
 
 The intense absorptions  at higher frequencies are due to CH stretches. The first peak at 
 2956 \cm (160  \AA$^4$/amu) is due to A$_1$ CH stretch where CH$_3$ belongs to 
the Si(CH$_3$)$_3$ group.  The A$_1$ CH stretch in the two methyls attached to silicon atoms leads
to absorption at slightly higher frequency  2969 \cm. The three-fold degenerate  asymmetric 
CH stretches absorb at  3032-3033 \cm ( 102 \AA$^4$/amu) and is also infra-red active.

 The most obvious difference between the Raman spectra of \SiA\, and \SilyA\, is the presence of 
the intense peak at 2130 \cm 
for the \SiA\,  involving SiH bonds and the high frequency modes involving CH bonds in the 
methylated \SilyA\, cluster. It may be mentioned here
that there are several very low intensity modes between 1000-1500 \cm for the \SilyA. 
However, the intensities of these peaks 
are comparatively very small and hence may not be used for identification purposes. 

   The passivation of dangling bonds at the silicon surfaces by hydrogenation, chlorination and alklation
 (ethylation, methylation etc) has been a subject of several studies due to technological importance of
 silicon.  Two groups have reported experimental studies of vibrational modes of methylated silicon 
 surface have been reported\cite{Webb}. Although the sila-adamantane core studied here is a very small unit of silicon
 crystal, it is interesting to compare the frequencies due to Si-C stretching and methyl deformations.
 Yamada and coworkers\cite{Yamada} assign the vibration of 678 \cm to Si-C which compare well with 664 \cm 
 Si-C asymmetric stretch.   The CH$_3$ stretch and deformation freuenqcies compare 
 well (Cf. \ref{tab:IR_big}).

\section{Conclusion}

     The equilibrium structure of recently synthesized  of silylated sila-adamantane  \silyA\,
was obtained within the density functional theory, using large polarized Gaussian 
basis sets and the semi-local PBE approximation to  the exchange-correlation functional. 
The calculated geometrical parameters are in very god agreement with experimental values.
The Si$_{10}$ core in the optimized \silyA\, is dilated with respected to the
silicon analogue of adamantane \siA. The calculation of vibrational frequencies 
indicate that some of the methyl ligands are freely rotating in the gas phase. 
The infra-red and  Raman spectra of \silyA\, is reported for the first time.  
The comparison of spectra for the  \silyA\,  with that of  \SiA\, do not show any 
similarity since in both the structures only the Si core is similar. The \SiA\, shows intense peaks corresponding to 
stretching of the Si-H bonds which are missing in the \SilyA\, 
due to absence of SiH bonds. On the other hand, the presence of methyls are
evident in the low frequency peaks for the \silyA\, 
and also in very high frequency peak arising from the CH bonds.

        This work is supported in part by the Office of Naval Research, directly and through the
 Naval Research Laboratory, the National Science Foundation through, and  partly by the 
University of Texas at El Paso (UTEP).  Authors acknowledge the computer time at the UTEP Cray 
acquired using ONR 05PR07548-00 grant.

\begin{thebibliography}{99}
\bibitem{adamantane}
P. von R Schleyer, J. Am. Chem. Soc. {\bf 79}, 3292 (1957); 
P. von R Schleyer, J. E. Williams, K. R. Blanchard, J. Am. Chem. Soc. {\bf 92}, 

\bibitem{Cheveron}
A. P. Marchand, Science, 299 (2003) 52;
J. E. P. Dahl, J. M. Moldowan, T. M. Peakman, J. C. Cardy, E. Lobkovsky, M. M. Olmstead, P. W. May, T. J. Davis, J. W. Steeds, K. E. Peters,
A. Pepper, A. Ekuan, and R. M. K. Carlson, Angew. Chem. Int. Ed. 42 (2003) 2040;
J. E. Dahl, S. G. Liu, and R. M. K. Carlson, Science  299 (2003) 96.

\bibitem{Diamondoids}
 M. Shen, H. F. Schaefer III, C. Liang, J. H. Lii, N. L. Allinger, P. von R. Schleyer,
 J. Am. Chem. Soc. {\bf 114}, 497 (1992);
 J. Y. Raty and G. Galli, Nat. Mater. {\bf 2}, 792 (2003); 
 S. L. Richardson, T. Baruah, M. J. Mehl, and M. R. Pederson, Chem. Phys. Lett. {\bf 403}, 83 (2005);
 J. Filik, J. N. Harvey, N. L. Allan, P. W. May, J. E. P. Dahl, S. Liu, and R. M. K. Carlson,
 Spectrochmica Acta Part A {\bf 64}, 681 (2005);
 Y.-R. Chen, H.-C. Chang, C.-L Cheng, C-C. Wang, and J. C. Jiang, J. Chem. Phys. 119 (2003) 10626;
 G. C. McIntosh, M. Yoon, S. Berber, and D. Tom\'{a}nek, Phys. Rev. B 70 (2004) 045401.


\bibitem{Expt}
  J. Fischer, J. Baumgartner, and C. Marschner, Science {\bf 310}, 825 (2005).

\bibitem{Pichierri} F. Pichierri, Chem. Phys. Lett. 421 (319 (2006).

\bibitem{KS} W. Kohn and L. J. Sham, Phys. Rev. 140 (1965) A1133.

\bibitem{HK}P. Hohenberg and W. Kohn, Phys. Rev.  136 (1964) B864.


\bibitem{PBE}
  J. P. Perdew, K. Burke, M. Ernzerhof,  Phys.  Rev.  Lett.   {\bf  77 }, 3865 (1996).


\bibitem{NRLMOL}
    M. R. Pederson, K. A. Jackson, Phys. Rev. B. {\bf 41}, 7453 (1990);
    K. A. Jackson, M. R. Pederson, Phys. Rev. B. {\bf 42}, 3276 (1990);
    M. R. Pederson, K. A. Jackson, Phys. Rev. B. {\bf 43}, 7312 (1991);
    M. R. Pederson, D. V. Porezag, J. Kortus, and  D. C. Patton,
      Phys. Stat. Sol. B 217 (2000) 197.


\bibitem{Porezag99}
D. V. Porezag, M. R. Pederson, Phys. Rev. A {\bf 60}, 2840 (1999).




\bibitem{ABINIT} 
X. Gonze, J.-M. Beuken, R. Caracas, F. Detraux, M. Fuchs, G.-M. Rignanese, 
L. Sindic, M. Verstraete, G. Zerah, F. Jollet, M. Torrent, A. Roy, 
M. Mikami,Ph. Ghosez, J.-Y. Raty, D.C. Allan.
Computational Materials Science {\bf 25}, 478-492 (2002).




\bibitem{benedict}
K. X. Benedict, A. Puzder, A. J. Williamsn, J. C. Grossman, G. Galli, J. Klepeis,
J. Raty, and O. Pankratov,  Phys. Rev. B {\bf 68}, 085310 (2003);
P. H. Han, W. G. Schmidt, and F. Becstedt, Phys. Rev. B {\bf 72}, 245425 (2005).


\bibitem{Yamada}
 T. Yamada, T. Inoue, K. Yamada, N. Takano, T. Osaka, H. Harada, K. Nishiyama, and I. Taniguchi, 
  J. Am. Chem. Soc. {\bf 125}, 8039 (2003).

\bibitem{Webb}
  L. J. Webb, S. Rvillion, D. H. Michalak, Y. J. Chabal, and N. S. Lewis, 
  J. Phys. Chem. B {\bf 110}, 7349 (2006).


\end{thebibliography}
\end{document}